\documentclass[prl,twocolumn,graphicx,amssymb,floatfix]{revtex4}

\begin{document}

{\bf Comment on ``How the result of a single coin toss can turn out to be 100 heads'' }

In a recent Letter,  Ferrie and  Combes \cite{FC} claimed to show ``that weak values are
not inherently quantum, but rather a purely statistical feature of pre- and post-selection with
disturbance.''  In this Comment I will show that this claim is not valid. It follows from  Ferrie and  Combes misunderstanding of  the concept of weak value.

Weak value of a variable $A$ is a  property of a single quantum system  pre-selected in a state $ |\psi \rangle$ and post-selected in a state $ |\phi \rangle$:
  \begin{equation}\label{wv}
  A_w\equiv\frac{\langle \phi |A|\psi\rangle}{\langle \phi |\psi\rangle}.
  \end{equation}
  Not only a measurement, but any coupling to a variable $A$ of the pre- and post-selected system, when it is weak enough, is an effective coupling to the weak value of this variable. In any such weak coupling, the operator $\hat A$ in the interaction Hamiltonian should be replaced by the c-number $A_w$. As a result, after the post-selection, the wave function of a quantum system weakly coupled to the observable is shifted (and renormalized, if the weak value is complex) in proportion to this weak value. 

Only when we want to observe this shift, we will need a pre- and post-selected ensemble and then the statistics becomes relevant. The weak value shifts exist if measured or not, so the  weak value is {\it not defined} by the statistics of measurement outcomes. The statistical analysis (performed after the post-selection) can just reveal the pre-existing weak values.

To prove their point, Ferrie and  Combes presented a purely classical situation with a coin toss which supposed to be analogous to the example presented in the first publication of the weak value which has the title: ``How the result of a measurement of a component of the spin of a spin-$\frac{1}{2}$ particle can turn out to be 100'' \cite{AAV}. However, I can see {\it nothing} in common. The weak value of a variable of a system is defined by pre-selected and post-selected states of the system. Weak value of 100 for spin $z$ component of a particle appeared for particular pre- and post-selected spin states:
\begin{eqnarray}
% \nonumber to remove numbering (before each equation)
 \nonumber  |\psi \rangle &=& \cos \frac{\alpha}{2}|\uparrow_x \rangle + \sin \frac{\alpha}{2}|\downarrow_x \rangle, ~~~~~\tan\frac{\alpha}{2}=100,\\
 |\phi \rangle &=& |\uparrow_x \rangle.
\end{eqnarray}
 Every weak enough coupling to the spin will show $(\sigma_z)_w=100$.
 In contrast, in  the examples of Ferrie and  Combes, the initial state is ``1'' and the final state \mbox{is ``-1''.} They got  a value   100   by playing with the definition of their  ``weak'' measurement. They could equally well get 1000. However, there is nothing in their construction analogous to (\ref{wv}) that  provides functional dependence on the pre- and post-selected  states of the system.

Failure to present a classical analog of the weak value measurement invalidates all  the conclusions of Ferrie and  Combes. The concept of weak value arises due to wave interference and has no analog in classical statistics.   Moreover, if weak values are observed with external systems (and not with a different degree of freedom of the observed system as it has been done until now) then the weak value appears due to interference of a quantum entangled wave  and it has no analog in classical wave interference too. Therefore, weak value is a genuinely quantum concept.

This work has been supported in part by the Israel Science Foundation  Grant No. 1311/14.

L. Vaidman\\
 Raymond and Beverly Sackler School of Physics and Astronomy\\
 Tel-Aviv University, Tel-Aviv 69978, Israel

\end{document}